\documentclass[aps,prl,twocolumn,superscriptaddress,showpacs]{revtex4}
\usepackage{graphicx}
\usepackage{mathrsfs}
\usepackage{bm}

\begin{document}
\title{Spin Polarized and Valley Helical Edge Modes in Graphene Nanoribbons}
\author{Zhenhua Qiao}
\affiliation{Department of Physics, The University of Texas, Austin,
Texas 78712, USA}
\author{Yugui Yao}
\affiliation{Institute of Physics, Chinese Academy of Sciences,
Beijing 100190, China} \affiliation{Department of Physics, The
University of Texas, Austin, Texas 78712, USA}
\author{Shengyuan A. Yang}
\affiliation{Department of Physics, The University of Texas, Austin,
Texas 78712, USA}
\author{Bin Wang}
\affiliation{Department of Physics, The University of Hong Kong,
Hong Kong, China}
\author{Qian Niu}
\affiliation{Department of Physics, The University of Texas, Austin,
Texas 78712, USA}

\begin{abstract}
Inspired by recent progress in fabricating precisely zigzag-edged
graphene nanoribbons and the observation of edge magnetism, we find
that spin polarized edge modes with well-defined valley index can
exist in a bulk energy gap opened by a staggered sublattice
potential such as that provided by a hexagonal Boron-Nitride
substrate. Our result is obtained by both tight-binding model and
first principles calculations. These edge modes are helical with
respect to the valley degree of freedom, and are robust against
scattering, as long as the disorder potential is smooth over atomic
scale, resulting from the protection of the large momentum
separation of the valleys.
\end{abstract}

\maketitle

\emph{Introduction.---} The appearance of edge states is one of the
most peculiar phenomena in solid state systems. They are often
connected to topologically non-trivial bulk properties, e.g.
non-zero Chern numbers in quantum Hall systems~\cite{TKNN,
Hatsugai}, or odd $Z_2$ numbers in time-reversal invariant
topological insulators~\cite{Kane2}. The edge states in quantum Hall
systems are robust against all kinds of disorders and
interactions~\cite{Niu1}, while those in the latter systems can
survive scatterings that preserve the time reversal
symmetry~\cite{Kane2,Kane1}. The edge states in graphene with zigzag
terminations belong to a different category. Such states connect the
two different valleys $K$ and $K^\prime$ projected along the edge
direction and their presence is dictated by the bulk topological
charge~\cite{WangYao}. It is of great interest to utilize these
unusual states for various applications~\cite{Louie,Kyle}.

Recently, zigzag-edged graphene nanoribbons have been fabricated
with precision by unzipping carbon
nanotubes~\cite{ZigzagCut,AnisotropicEtching,ZigzagReconstruct}.
Without electron-electron interaction, the edge states form a
completely flat edge band connecting the two valleys with large
momentum separation~\cite{ExpGraphene,RMPGraphene}. When interaction
is taken into account, due to the singular density of states, spins
on the edge become spontaneously polarized resulting in an edge
ferromagnetism ~\cite{Louie,ZigMagAbinitio}, which has been
confirmed by a recent experiment~\cite{ZigMagExp}. The spin
polarized edge states are dispersive in momentum space, making them
useful for current transport. Unfortunately, without a bulk gap, the
effect of edge states would be overwhelmed by the contribution from
the bulk states.

In this Letter, we show that for zigzag-edged graphene nanoribbons
spin polarized dispersive edge states can exist and remain robust in
a bulk gap opened by a staggered sublattice potential. Such
potential can be realized, for example, by a hexagonal Boron-Nitride
substrate. The edge modes are then tied to the conduction or valance
band edges, and with spontaneous spin polarization due to
interaction effects, one spin branch of the edge modes is pushed
into the bulk gap forming a conducting channel in the bulk
insulating state. We show that such spin-polarized states within the
bulk gap remain robust under scattering with correlation length
longer than the lattice constant. We also perform first principles
calculations to further support our predictions.

\begin{figure}
\includegraphics[width=8cm,totalheight=5cm,angle=0]{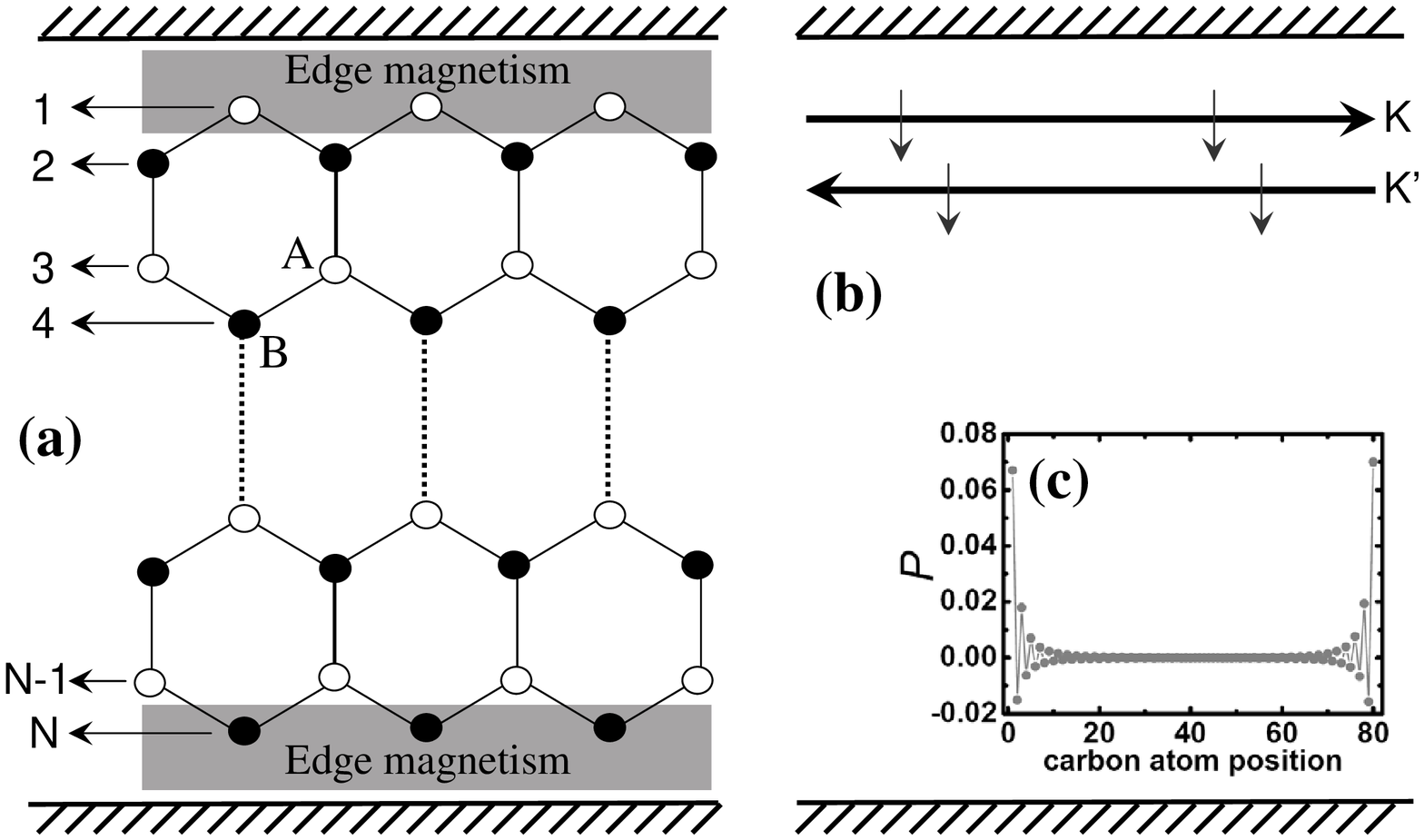}
\caption{(a) Diagram of zigzag-edged graphene nanoribbons. A
($\circ$) and B ($\bullet$) sublattices are subjected to staggered
potentials. Magnetization $M_0$ only exists at the edge atoms in the
grey regime. $N$ labels the ribbon width. (b) Schematic diagram of
edge states propagation in a two-terminal measurement geometry for
Fermi energy in the bulk gap of Figs.~\ref{fig2}(b)-(d). (c) First
principles calculation of the spin-polarization $P$ for the carbon
atoms of the supercell of zigzag graphene on top of the hexagonal
Boron Nitride substrate.
$P=(n_\uparrow-n_\downarrow)/(n_\uparrow+n_\downarrow)$.}
\label{fig1}
\end{figure}

\emph{Tight-Binding Model.---} Figure \ref{fig1}.(a) illustrates the
schematic setup of a zigzag-edged graphene nanoribbon in the
presence of a staggered sublattice potential. A tight-binding
Hamiltonian that incorporates phenomenologically the edge
spin-polarization can be written as:
\begin{equation}
H=-t \sum_{\langle{ij}\rangle}{ c^\dagger_{i}c_{j}} +
M_0{\sum_{i=1,N}} { c^\dagger_{i} \sigma_{z} c_{i}}+\sum_{i}{U_{i}
c^\dagger_{i} c_{i}}
\end{equation}
where $c^\dag_{i}$($c_{i}$) is the electron creation (annihilation)
operator on site $i$, and $\sigma_z$ is the $z$ component of Pauli
matrices. The first term is the nearest neighbor hopping with $t$
being the hopping energy. The second term represents the effect of
edge ferromagnetism which stems from electron-electron interaction,
and $M_0$ is a phenomenological parameter put in by hand at present
stage, and its value will be determined later from the
first-principles method. The last term corresponds to the staggered
$A/B$ sublattice potentials: $U_i$=$\Delta/2$ for sublattice A
($\circ$), and $U_i$$=$$-\Delta/2$ for sublattice B ($\bullet$). In
our analysis of tight-binding model, we measure the energy
$\varepsilon$, magnetization $M_0$, potential $\Delta$, and disorder
strength $W$ in units of the hopping energy $t$.

\begin{figure}
\includegraphics[width=8.5cm,totalheight=6.cm,angle=0]{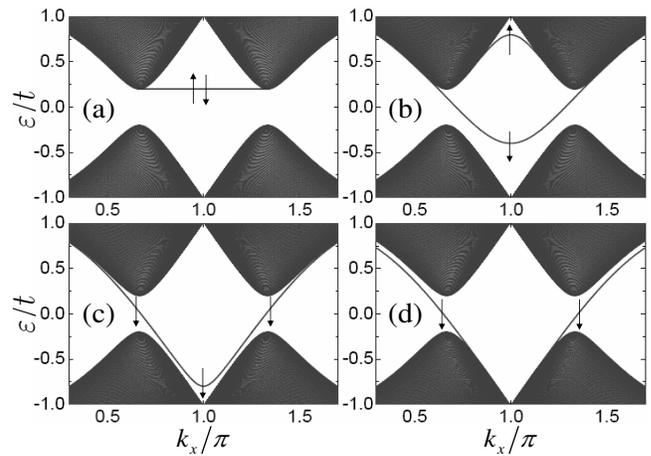}
\caption{Evolution of the band structures of zigzag edged graphene
nanoribbons at fixed width $N=800$~(only exhibiting the edge states
from the upper boundary). (a) When the staggered sublattice
potential $\Delta=0.4$ is applied, an bulk gap is opened, and the
flat-bands are doubly-degenerate; (b)-(d) The edge magnetism is
further included with $M_0=0.6,~1.0,$ and 1.4, respectively. The
flat bands become spin-split: spin-up edge band bends upwards, while
spin-down edge band bends downwards.} \label{fig2}
\end{figure}

The energy spectrum of graphene nanoribbons with zigzag termination
can be numerically obtained by diagonalizing the Hamiltonian
$H(\bm{k})$ in momentum space for each crystal momentum $\bm{k}$
along the edge direction which we take as $x$-axis.
Figure~\ref{fig2} shows the evolution of the energy spectrum as
functions of staggered sublattice potential $\Delta/2$ and edge
magnetization $M_0$ for fixed ribbon width $N=800$ (about 852 {\rm
\AA}). For clarity, only the edge states from the upper boundary are
shown. Panel (a) shows the doubly-degenerate flat-bands connecting
the two Dirac points $K$ and $K'$. We observe that a bulk energy gap
$\Delta=0.4$ is opened by inversion symmetry breaking due to the
staggered sublattice potentials.

Figures~\ref{fig2}.(b)-(d) plot the band structures when $M_0$ is taken to be $0.6$, $1.0$, and $1.4$,
respectively. We find that, due to the different degrees of
localization of the states in edge band~\cite{WangYao}, the magnitude
of the energy splitting of the edge bands is $k$-dependent: the spin-up edge
band bends upward, while the spin-down edge band bends downward. This
makes edge band dispersive hence capable of conducting charge current. Moreover, for
a fixed sublattice potential $\Delta/2$$=$$0.2$, we find that along
with the increasing of the edge magnetization $M_0$ from 0.6 [see
panel (b)] to 1.0 [see panel (c)], the spin-down edge band gradually
approaches the bulk valence band, and eventually touches and
emerges into the bulk valence band (at $M_0$$\simeq$$1.4$). This creates gapless
edge modes tied to each valley, which is similar to the finding in
Ref.~\cite{WangYao} except that the edge modes here are spin-polarized.

From the energy spectra, we observe that the two edge states in the
bulk energy gap propagate along opposite directions. Due to the same
flat-band origin, they are localized at the same upper boundary. We
also note that the edge states in the gap have well-defined valley
index $K$ or $K^\prime$. The situation is schematically shown in
Fig.~\ref{fig1}(b): the edge states with same spin but different
valley indices propagate oppositely along the same boundary. This
can be natually termed as a spin-polarized quantum-valley Hall
state.

\emph{Robustness of Spin-Polarized Edge Mode.---} From the above
analysis, we notice that for the fixed bulk gap, the spin polarized
edge state is gapped for weak edge magnetization, while the edge
state becomes gapless when the edge magnetization approaches a
critical $M^c_0$. These edge states in the gap provide conducting
channels for spin-polarized transport. However, to be useful for
practical applications, they need to be robust against impurity
scattering. In the following, we shall investigate the robustness of
the edge state in the presence of impurities, and show that both
gapped and gapless edge states are robust against scattering due to
the large momentum separation between the valleys $K$ and
$K^\prime$.

It is known that the impurity scattering in graphene mainly comes
from the long-range Coulomb scatterers~\cite{LongRangeDisorder}. We
assume that the impurity potential $V_i$ at each site $i$ takes a
Gaussian form~\cite{DisorderForm}:
\begin{eqnarray}
V_i=\sum_{j} w_j \times \exp
(-\frac{|\textbf{r}_j-\textbf{r}_i|^2}{2\times \xi^2})
\end{eqnarray}
where the summation is over all sites, $w_j$ is the local disorder
strength at site $j$, and $\xi$ is the correlation length. We define
an effective disorder strength $W$ from $\xi$ and
$w$~\cite{definition}:
\begin{eqnarray}
W= w \times ({\xi} ^2 +1)
\end{eqnarray}
The numerical simulations are performed within the same setup of
Ref.~\cite{QuantumSpinHallEffect} including only the left and right
semi-infinite leads. The two-terminal conductance can be calculated
from the Landauer-B\"{u}ttiker formula~\cite{datta}:
\begin{eqnarray}
G=\frac{e^2}{h} \rm{Tr}[{\Gamma_R G^r \Gamma_L G^a}]
\end{eqnarray}
where $G^{r,a}$ are the retarded and advanced Green's functions of
the central disordered region. The quantities ${\Gamma_{L/R}}$ are
the line-width functions describing the coupling between the
left/right lead and the scattering region, and can be obtained from
${\Gamma_{P}}=i (\Sigma^r_{P}-\Sigma^a_{P})$. Here,
$\Sigma^{r/a}_{P}$ are the retarded/advanced self-energies of the
$P-th$ semi-infinite lead, and can be numerically evaluated using
the recursive transfer matrix method~\cite{selfenergy}.

\begin{figure}
\includegraphics[width=8.5cm,totalheight=5.8cm,angle=0]{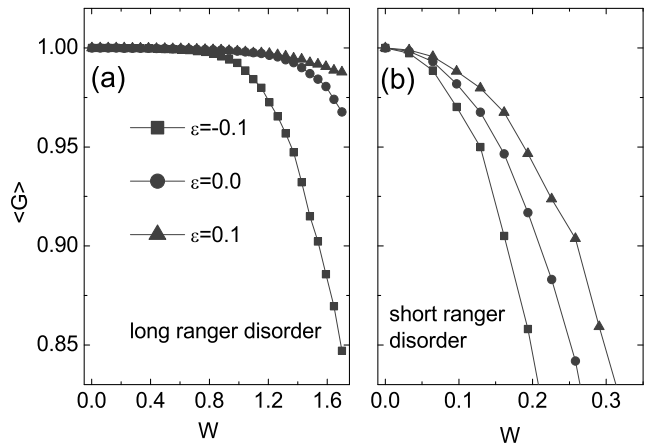}
\caption{Averaged conductance $\langle G \rangle$ (in units of
$e^2/h$) versus effective disorder strength $W$ at fixed
$\Delta$=0.4, and $M_0$=0.6 for different Fermi energies
$\varepsilon=-0.1$(square),~0(circle),~0.1(triangle), respectively.
(a) Long ranger disorder ($\xi$=4); (b) Short ranger disorder
($\xi$=0). 20~000 ensembles are collected for each set of
parameters.} \label{fig3}
\end{figure}

Figure \ref{fig3} plots the sample-averaged two-terminal conductance
$\langle G \rangle$ as a function of the effective disorder strength
$W$ for three different Fermi energies $\varepsilon$=-0.1,~0,~0.1,
respectively. The edge magnetization is set to be $M_0$=0.6. Each
data point represents average over 20,000 sample configurations.
Panel (a) is for the long range disorder case. We observe that, for
all the three Fermi energies, the averaged conductances $\langle G
\rangle$ are robust against weak disorders, i.e. when $W$$<$0.8, all
the conductances are exactly quantized to be one in units of $e^2/h$
without conductance fluctuation. When $W$$>$0.8, we find that
$\langle G \rangle$ of $\varepsilon$=-0.1 is quickly destroyed
first, and that of $\varepsilon$=0.1 is the most robust one. This
can be explained from the band structure shown in
Fig.~\ref{fig2}(b). One can see that the two edge states for a fixed
Fermi energy have a large momentum separation when the Fermi energy
is near the upper band bottom (e.g. $\varepsilon$=0.1). The
separation decreases when the Fermi energy is approaching the
valence band top (e.g. $\varepsilon$=-0.1). The large momentum
separation (on the scale of valley separation) suppresses the long
range impurity scattering which allows only small momentum transfer.
Panel (b) shows the averaged conductance as a function of the short
range nonmagnetic disorders, with other parameters being the same as
that in panel (a). We find that the edge states are very sensitive
to the disorders and therefore easily destroyed by small disorder
strengths. We also performed calculations for magnetic disorders
(not shown here) and the results are similar. Therefore, we conclude
that our valley associated spin-polarized edge modes are robust
against smooth disorder scattering.

\begin{figure}
\includegraphics[width=8.cm,totalheight=10cm,angle=0]{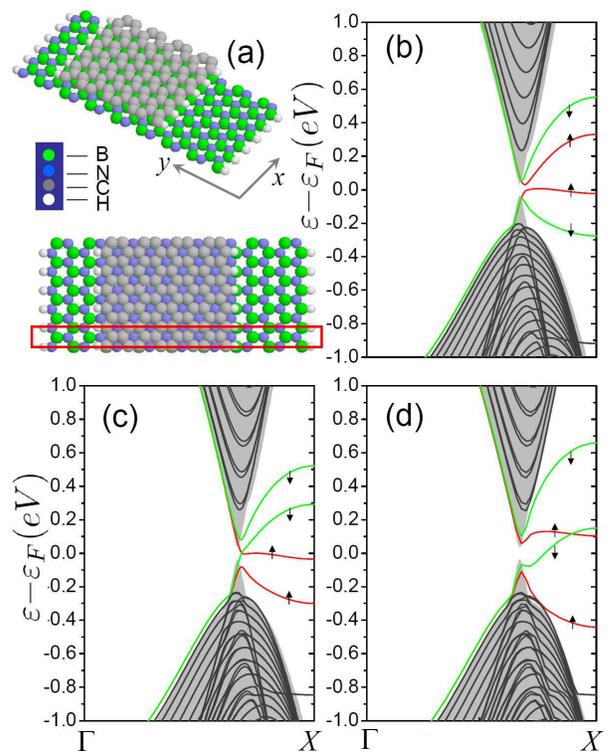}
\caption{(Color online) (a) Atomic structure of hydrogen-terminated
zigzag-edged graphene nanoribbons on top of single layer hexagonal
Boron-Nitride. The red square regime represents a supercell. Upper:
side view; lower: top view. (b) Band structure for spin
anti-parallel configurations between two zigzag edges. In the bulk
energy gap $\Delta$ around $78meV$ (the narrow energy window of the
projected bulk band structure shown in light grey), only spin-up
states exist, and a trivial gap is opened due to the
electron-electron interaction with the same spin between the two
zigzag edge boundaries. (c) Band structure for spin parallel case.
Spin-up and spin-down states coexist in the gap. (d) When a weak
voltage bias $0.27V$ is applied transversely, the upper (lower) edge
states are upwards (downwards) lift, leaving only the spin-down
states in the gap. Bands in Red and green represent spin-up and spin
down edge bands, respectively.} \label{fig4}
\end{figure}

\emph{First Principles Calculations.---} So far, we have
investigated the spin-polarized edge modes in a graphene nanoribbon
model Hamiltonian with a staggered $AB$ sublattice potential, and an
edge-specific spin polarization put in by hand. In the following,
from first principles calculations, we provide a concrete system for
the realization of our models by placing a zigzag-edged graphene
nanoribbon on top of a hexagonal Boron-Nitride substrate. In our
calculations, we set the lattice constant to be $a=2.45\rm{\AA}$,
and inter-layer distance $d=3.22 \rm{\AA}$~\cite{AbinitioB6N6}.
Figure \ref{fig4}.(a) illustrates the schematic configuration of the
system. Here, we use $N_1$ ($N_2$) to label the width of graphene
(Boron nitride), and $N_1$$<$$N_2$. The single-layer graphene and
Boron nitride are $AB$ stacked with Nitrogen atoms on top of the
hollow position. All the outer-most atoms are saturated with
Hydrogen atoms, and we use the experimental values of the bond
lengths: $1.17\rm{\AA}$~(B-H), $1.01\rm{\AA}$~(N-H), and
$1.09\rm{\AA}$~(C-H). The self-consistent ground state calculations
were performed within the non-equilibrium Green's function coupled
with the density-functional theory scheme~\cite{Jeremy}, and the
local density approximation exchange-correlation potential
(LDA-PZ81) was used\cite{LDA_PZ81}.

In our calculations, we set $N_1=96$, and $N_2=112$. Panels (b)
and (c) plot the band structures with spin anti-parallel/parallel
configurations at the two zigzag boundaries. The grey region shows
bulk band structure region (when both $N_1$ and $N_2$ approach
infinity): a bulk gap $\Delta$ around $78~meV$ is opened, which is
slightly larger than $53~meV$ using VASP package\cite{AbinitioB6N6}.
In panel (b), we find that only the spin-up polarized edge states
lie inside gap, which resembles the band structure of the
tight-binding model. A small splitting $\delta$ in the figure
is due to the interaction between the states
with the same spin on the two boundaries, and will decrease along with the
increasing system width. In panel (c), we observe that
both the spin-up and spin-down states coexist in the gap. From both
panels (b) and (c), one can obtain that the magnetization of each
outermost carbon atom is about $M_0$=$0.287~eV$. As shown in panel
(d), one can apply an external transverse bias $V=0.27~V$ across the
ribbons~\cite{Louie} to separate the mixed states and leave only the
spin-down edge state in the bulk gap, which provides an efficient way to
manipulate the spin-polarized edge states.

\emph{Conclusion.---} We investigate on the edge modes of
zigzag-edged graphene nanoribbons in the presence of a staggered
sublattice potential. We find that the edge states form
spin-polarized conducting channels which are robust against smooth
impurity potentials. Using first principles calculation methods, we
provide a specific system which exhibits such spin-polarized edge
modes by placing the zigzag-edged graphene nanoribbons on top of a
hexagonal Boron-Nitride substrate. The realization of this valley
associated spin-polarized edge modes will enable the application of
the graphene-based spintronics and valleytronics devices.

Z.Q. was supported by NSF~(DMR0906025) and Welch
Foundation~(F-1255). Q.N. was supported by DOE~(DE-FG02-02ER45958,
Division of Materials Science and Engineering) and Texas Advanced
Research Program. Y.Y. was supported by NSF of China (Grants
No.~10974231) and the MOST Project of China (Grants
No.~2007CB925000, and 2011CBA00100).


\begin{thebibliography}{99}
\bibitem{definition}
This can be obtained through the Fourier transformation, and we have
further numerically verified it.
\end{thebibliography}
\end{document}